 \documentclass[twocolumn,showpacs,amsmath,amssymb,prc]{revtex4}

\usepackage{CJK} 

\usepackage{graphicx,amsmath,amssymb,bm,multirow}
\usepackage{amsthm}
\usepackage{amsfonts}
\usepackage{dcolumn}
\usepackage{bm}
\usepackage[colorlinks,linkcolor=blue,anchorcolor=blue,citecolor=blue]{hyperref}
\usepackage[usenames,dvipsnames]{color}

\newcommand{\beq}{\vspace{0.5em}\begin{equation}}
\newcommand{\eeq}{\end{equation}\vspace{0.5em}}
\newcommand{\beqn}{\vspace{0.5em}\begin{eqnarray}}
\newcommand{\eeqn}{\end{eqnarray}\par\vspace{0.5em}\noindent}

\newcommand{\bsub}{\begin{subequations}}
\newcommand{\esub}{\end{subequations}}

\begin{document}
%

\begin{CJK*}{GB}{song}

\title{Beyond mean-field approach for pear-shaped hypernuclei}

\author{HaoJie Xia} 
\affiliation{School of Mechanical and Electrical Engineering, Handan University, Handan  056005, Hebei, China}

\author{XianYe Wu} 
\affiliation{College of Physics and Communication Electronics, Jiangxi Normal University, Nanchang 330022, China} 

\author{Hua Mei} 
\affiliation{Department of Physics and Astronomy, University of North Carolina, Chapel Hill, North Carolina 27516-3255, USA}
\affiliation{Department of Physics, Tohoku University, Sendai 980-8578, Japan} 


\author{JiangMing Yao}\email[]{Corresponding author: yaoj@frib.msu.edu}
\affiliation{FRIB/NSCL, Michigan State University, East Lansing, Michigan 48824, USA}  

\begin{abstract} 

We develop both relativistic mean field and beyond approaches for hypernuclei with possible  quadrupole-octupole deformation or  pear-like shapes based on relativistic point-coupling energy density functionals.  The symmetries broken in the mean-field states are recovered with parity, particle-number and angular momentum projections. We take $^{21}_\Lambda$Ne as an example to illustrate the method, where the $\Lambda$ hyperon is put on one of the two lowest-energy orbits (labeled as $\Lambda_s,  \Lambda_p$), respectively.  We find that the $\Lambda$ hyperon in both cases disfavors the formation of a reflection-asymmetric molecular-like $^{16}$O$+\alpha$ structure in $^{20}$Ne, which is consistent with the Nilsson diagram for the hyperon in $(\beta_2, \beta_3)$ deformation plane. In particular, we show that the negative-parity states with the configuration $^{20}$Ne($K^\pi=0^-)\otimes \Lambda_s$ are close in energy to those with the  configuration $^{20}$Ne($K^\pi=0^+)\otimes \Lambda_p$, even though they have very different structures. The $\Lambda_s$ ($\Lambda_p$)  becomes more and more concentrated around the bottom (top) of the ``pear" with the increase of octupole deformation.
\end{abstract}

\pacs{21.60.Jz, 21.80.+a, 27.30.+t} 

\maketitle

 \section{\label{introduction} Introduction}

Hypernuclei provide a natural and unique laboratory to study hyperon-nucleon interaction in nuclear matter, the knowledge of which is relevant for understanding the structure of neutron stars~\cite{Glendenning00}.  Hypernuclei are self-bound quantum many-body systems composed of three different types of fermions. In  single-$\Lambda$ hypernuclei, the unpaired $\Lambda$ hyperon is free of Pauli-exclusion principle from nucleons and thus can occupy any energy-allowed orbit, generating many hypernuclear states. Some of them are so-called {\em genuine} hypernuclear states which do not have corresponding states in  ordinary nuclei \cite{Motoba83}. This unique character makes it interesting to study the structure of hypernuclei and the hyperon impurity effect in atomic nuclei.

The self-consistent mean-field approaches starting from effective interactions or universal energy density functionals (EDFs) with their parameters determined directly from the properties of nuclear many-body systems have achieved a great success in the studies of both ordinary nuclei and hypernuclei ranging from light to heavy mass regions for several decades. The relativistic mean-field (RMF)  approaches \cite{Ring96,Vretenar05,Meng06} are of particular interest in nuclear physics as Lorentz invariance is one of the underlying symmetries of QCD. This symmetry not only allows to describe the spin-orbit coupling of nucleons, which has an essential influence on the underlying shell structure, in a consistent way, but also puts stringent restrictions on the number of parameters in the corresponding functionals. Based on the picture of exchanging some effective mesons to simulate the features of $NN$  and $\Lambda N$ interactions in the nuclear medium, the RMF approaches have been extensively adopted for studying the $\Lambda$ binding energies of  spherical hypernuclei. The major findings in these studies include the shrinkage effect of the  $s$-orbital $\Lambda$ hyperon and the important role of tensor coupling between the vector $\omega$ meson  and the $\Lambda$ hyperon in reproducing a weak $\Lambda$ spin-orbit interaction. See a review~\cite{Hagino16}  and references therein for the studies of hypernuclei in the RMF approaches.

In the recent decade, the RMF study of $\Lambda$ hypernuclei has been extended to deformed $\Lambda$ hypernuclei, including axially~\cite{Win08,Lu11,Xu15} and triaxially deformed~\cite{Sang13,Xue15}  shapes.  The  mean-field calculation based on a Skyrme EDF has shown that the $\Lambda$ hypernucleus shares a similar deformed shape as that of the corresponding core nucleus ~\cite{Zhou07}. This finding was confirmed in the axially deformed RMF calculation~\cite{Win08}, where the authors have also disclosed some exceptions, such as $^{13}_\Lambda$C and $^{29}_\Lambda$Si, for which, the equilibrium shape is changed from an oblate shape to the spherical one with the
presence of a $\Lambda$ in the lowest-energy state (labeled as $\Lambda_s$). The significant shape polarization of the $\Lambda$ hyperon has been confirmed in two subsequent RMF calculations based on either the meson-exchange $\Lambda N$ interaction~\cite{Lu11} or the contact $\Lambda N$ interaction~\cite{Xue15}, even though the mean-field approximation is prone to overestimate the hyperon impurity effect on nuclear shapes \cite{Mei18}. 
Moreover, it has been found in the calculations of both the RMF~\cite{Xue15} and the  antisymmetrized molecular dynamics (AMD)~\cite{Isaka13}  models  that the deformation of the hypernucleus can also be enlarged if the $\Lambda$ occupies the excited states with the orbital angular momentum $\ell \ge1$. The shape polarization effects of the $\Lambda$ in different orbits can be understood from its Nilsson diagram. Generally, the energy of the $\Lambda_s$ increases with the quadrupole deformation and thus it becomes less bound in a deformed hypernuclear state~\cite{Xue15} and an opposite behavior is found for the $\Lambda$ in the second lowest-energy state (labeled as $\Lambda_p$). This argument is, however, not necessarily true when an exotic structure appears in the deformed state, as found for example in the normal and superdeformed states of $^{37}_\Lambda$Ar~\cite{Lu14,Wu17}.

Significant progress has been achieved in the beyond mean-field studies of hypernuclei  in recent years. The beyond mean-field approaches implemented with projection techniques and generator coordinate method (GCM) are based on either the Gogny force \cite{Isaka11AMD,Isaka11Ne,Isaka12},  Skyrme EDF \cite{Cui15,Cui17}, or relativistic EDF  \cite{Mei14,Mei15,Mei16,Mei17,Mei16R} for the effective nucleon-nucleon interactions, together with different types of effective hyperon-nucleon interactions. These approaches provide useful tools to analyze the spectroscopy of hypernuclear low-lying states and build connections between the hypernuclear spectroscopic data and the underlying effective interactions.  It has been found that the low-lying positive-parity states of even-even-one hypernuclei can be well described with the single configuration $[^{A-1}Z (I^+) \otimes \Lambda s_{1/2}]^J$. In contrast,  there is a large configuration mixing between the two components $[^{A-1}Z (I^+) \otimes \Lambda p_{1/2}]^J$  and $[^{A-1}Z (I\pm2 ^+) \otimes \Lambda p_{3/2}]^J$  in the hypernuclear $1/2^-_1, 3/2^-_1$ states, as illustrated in both samarium \cite{Mei17} and carbon \cite{xia2017}  hypernuclei.  The mixing weight increases as the collective correlation of nuclear core becomes stronger.  Based on the nature of these two states,  it was suggested that the spin-orbit  splitting size of $p_\Lambda$  hyperon state might be overestimated using the data of  $^{13}_\Lambda$C. Moreover, $^{15}_\Lambda$C was suggested to be the best candidate among the carbon hypernuclei to study the spin-orbit splitting of $p_\Lambda$  hyperon state \cite{xia2017}.

Recently, the beyond RMF approach with parity, particle-number and angular momentum projections was developed for the low-lying parity doublets of octupole shaped ordinary nuclei~\cite{Yao15,Zhou16}. In the mean time, this framework was extended to the low-lying states of hypernucleus $^{21}_\Lambda$Ne~\cite{Mei16R}, where the octupole shape degree of freedom was, however, not taken into account.  In this work, we combine the virtues of the above two methods and develop a beyond RMF approach for the low-lying states of octupole shaped $\Lambda$ hypernuclei. 
 
 For the hypernucleus with a nonzero octupole deformation, parity violation is allowed in the single-particle state of the $\Lambda$.  As a result, the wave functions of both $\Lambda_s$ and $\Lambda_p$ are expected to be dominated by the admixtures of  $s$ orbit and  $p$ orbit. It is interesting to know how the $\Lambda_s, \Lambda_p$ change the topology of the energy surface and how the octupole correlation changes the low-lying states of hypernuclei.  $^{20}$Ne provides a good example for this purpose as the low-lying negative-parity band $K^\pi=0^-$ built on the configurations with the reflection-asymmetric molecular structure $^{16}$O$+\alpha$  is observed. The band-head energy is around 5.8 MeV, which is comparable to the typical energy to excite a $\Lambda_s$  to $\Lambda_p$. The presence of non-zero octupole deformation brings the so-called ``parity-coupling" effect in $^{21}_\Lambda$Ne, which has been studied very early with a three-cluster model~\cite{Yamada84} and later with a microscopic cluster model~\cite{Hiyama09} and the hypernuclear AMD model~\cite{Isaka11}. In contrast to the hypernuclear AMD model,  our method is not necessarily limited to light hypernuclei, but can also be applied for heavier systems. Moreover, the excitation state of the hyperon can be easily treated in this method. As an illustration of the method,  we will focus on only the configurations $^{20}$Ne($K^\pi=0^-)\otimes \Lambda_s$ and $^{20}$Ne($K^\pi=0^+)\otimes \Lambda_p$  separately in this work for the sake of simplicity.

The paper is arranged as follows. In Sec.~\ref{Sec.II}, we present the main formalism of the beyond
RMF approach for octupole shaped $\Lambda$ hypernuclei. In Sec.~\ref{Sec.III}, we present  the results 
from both mean-field and projection calculations. 
A summary of the present study and an outlook are then given in Sec.~\ref{Sec.IV}.
 
 \section{The model}%
 \label{Sec.II}

\subsection{The RMF approach for $\Lambda$ hypernuclei}
The Lagrangian density ${\cal L}$ for a $\Lambda$ hypernucleus in the RMF approach  based on  pointing-coupling effective interactions can be written as 
  \begin{equation}
  \label{Lag1}
  {\cal L} = {\cal L}^{\rm free} + {\cal L}^{\rm em}    + {\cal L}^{NN} + {\cal L}^{N\Lambda},
   \end{equation}
  where the Lagrangian density   ${\cal L}^{\rm free}$ for free baryons and that ${\cal L}^{\rm em}$ for the electromagnetic field  are
  \begin{eqnarray}
    {\cal L}^{\rm free} &=& \sum_{B=N, \Lambda}\bar\psi^B ( i\gamma^\mu\partial_\mu - m_B)\psi^B, \\
    {\cal L}^{\rm em} &=& -\frac{1}{4}F^{\mu\nu} F_{\mu\nu} - e\bar\psi\gamma_\mu\frac{1-\tau_3}{2}\psi  A^\mu.
 \end{eqnarray}
 The $\psi^B$ represents either the nucleon ($B=N$) or hyperon ($B=Y$) field,  $m_B$ for the corresponding mass and $F^{\mu\nu}$ for the field tensor of the electromagnetic field $A^\mu$, defined as $F^{\mu\nu}=\partial^\mu A^\nu- \partial^\nu A^\mu$.  The  ${\cal L}^{NN}$ term is for the effective $NN$ interaction~\cite{Burvenich2002_PRC65-044308,Zhao10}
\begin{eqnarray}
\label{Lagrangian-NN}
&& {\cal L}^{NN}  \nonumber\\ 
   &=&  -\frac{1}{2}\alpha_S(\bar\psi^N\psi^N)^2
                -\frac{1}{2}\alpha_{TS}(\bar\psi^N\vec\tau\psi^N)^2\nonumber\\
             && -\frac{1}{2}\alpha_{V}(\bar\psi^N\gamma_\mu\psi^N)^2
                             -\frac{1}{2}\alpha_{TV}(\bar\psi^N\vec\tau\gamma_\mu\psi^N)^2\nonumber\\
                &&-\frac{1}{3}\beta_S(\bar\psi^N\psi^N)^3-\frac{1}{4}\gamma_S(\bar\psi^N\psi^N)^4\nonumber\\
                &&
                   -\frac{1}{4}\gamma_V[(\bar\psi^N\gamma_\mu\psi^N)(\bar\psi^N\gamma^\mu\psi^N)]^2 \nonumber\\
                &&   -\frac{1}{2}\delta_S\partial_\nu(\bar\psi^N\psi^N)\partial^\nu(\bar\psi^N\psi^N)\nonumber\\
                &&
                -\frac{1}{2}\delta_{TS}\partial_\nu(\bar\psi^N\vec\tau\psi^N)\cdot\partial^\nu(\bar\psi^N\vec\tau\psi^N)\nonumber\\
             &&
             -\frac{1}{2}\delta_V\partial_\nu(\bar\psi^N\gamma_\mu\psi^N)\partial^\nu(\bar\psi^N\gamma^\mu\psi^N)\nonumber\\
             &&
             -\frac{1}{2}\delta_{TV}\partial_\nu(\bar\psi^N\vec\tau\gamma_\mu\psi^N)\cdot\partial^\nu(\bar\psi^N\vec\tau\gamma^\mu\psi^N),  
\end{eqnarray}
 and  ${\cal L}^{N\Lambda}$  for the $N\Lambda$  interaction~\cite{Tanimura2012_PRC85-014306}
\begin{eqnarray} 
{\cal L}^{N\! \Lambda}&=& -\alpha_S^{(N\!\Lambda)}(\bar{\psi}^{N}\psi^N)(\bar{\psi}^{\Lambda}\psi^{\Lambda})\nonumber\\ 
&&
-\alpha_V^{(N\!\Lambda)}(\bar{\psi}^{N}\gamma_{\mu}\psi^N)
(\bar{\psi}^{\Lambda}\gamma^{\mu}\psi^{\Lambda})\nonumber\\ 
&&-\delta_S^{(N\!\Lambda)}(\partial_{\mu}\bar{\psi}^{N}\psi^N)
(\partial^{\mu}\bar{\psi}^{\Lambda}\psi^{\Lambda}) \nonumber\\
&& -\delta_V^{(N\!\Lambda)}(\partial_{\mu}\bar{\psi}^{N}\gamma_{\nu}\psi^N)
(\partial^{\mu}\bar{\psi}^{\Lambda}\gamma^{\nu}\psi^{\Lambda})\nonumber\\
 &&+
\alpha^{(N\!\Lambda)}_T(\bar{\psi}^{\Lambda}\sigma^{\mu\nu}\psi^{\Lambda})
(\partial_{\nu}\bar{\psi}^{N}\gamma_{\mu}\psi^N).
\end{eqnarray} 

The EDF derived from the above Lagrangian density can be separated into two parts
\begin{eqnarray}
E^{N}_{\textrm{RMF}}  &=& T_N + \int d^3r\varepsilon_{NN}(\textbf{r})+ \frac{1}{2}A_0 e\rho_V^{(p)}, \\
E^{\Lambda}_{\textrm{RMF}}
&=& T_\Lambda +\int d^3r\varepsilon_{N\Lambda}(\textbf{r}),
\end{eqnarray}
where the first term $T_{B=N/\Lambda}={\rm Tr}[(\vec{\alpha}\cdot\vec{p}+m_B\beta)\rho^B_V]$ is for the kinetic energy of nucleons or $\Lambda$ hyperon. The interaction energy terms are as follows
\begin{eqnarray}\nonumber
\label{Energy:Nh}
\varepsilon_{NN}&=&\frac{1}{3}\beta_S(\rho_S^N)^3+\frac{1}{4}\gamma_S(\rho_S^N)^4 +\frac{1}{4}\gamma_V(\rho_V^N)^4\\
&&+\frac{1}{2}\sum\limits_K\alpha_K(\rho_K^N)^2+\frac{1}{2}\sum\limits_{K=S,V,TV}\delta_K\rho_K\Delta\rho_K^N\\\nonumber
 \varepsilon_{N\Lambda}&=&\sum\limits_{K=S,V}\alpha_K^{(N\Lambda)}\rho_K^N\rho_K^\Lambda+\sum\limits_{K=S,V}\delta_S^{(N\Lambda)}\rho_K^N\Delta\rho_K^\Lambda\\
 &&+\alpha_T^{(N\Lambda)}\rho_V^N\rho_T^\Lambda,
\end{eqnarray}
where the densities are defined as
\begin{eqnarray}
\rho^N_S&=&\sum \limits_k\bar\psi^N_k\psi^N_k , \hspace{1cm}
\rho^N_V=\sum \limits_k\psi^{N\dagger}_k\psi^N_k ,\hspace{1cm}\\
\rho^N_{TS}&=&\sum\limits_k\bar\psi^N_k\tau_3\psi^N_k ,\hspace{0.5cm}
\rho^N_{TV}=\sum\limits_k\psi^{N\dagger}_k\tau_3\psi^N_k,\\
\rho^\Lambda_S&=&\sum \limits_k\bar\psi^\Lambda_k\psi^\Lambda_k,\hspace{1cm}
\rho^\Lambda_V=\sum \limits_k\psi^{^\Lambda\dagger}_k\psi^\Lambda_k ,\\
\rho^\Lambda_T&=&\nabla\cdot(\bar\psi_\Lambda i\vec{\alpha}\psi_\Lambda).
\end{eqnarray}
 It contains sixteen coupling constants $\alpha_S$, $\alpha_V$, $\alpha_{TS}$, $\alpha_{TV}$, $\alpha_S^{(N\Lambda)}$, $\alpha_V^{(N\Lambda)}$, $\alpha_T^{(N\Lambda)}$, $\beta_S$, $\gamma_S$, $\gamma_V$, $\delta_S$, $\delta_V$, $\delta_{TS}$, $\delta_{TV}$, $\delta_S^{(N\Lambda)}$ and $\delta_V^{(N\Lambda)}$.  The subscript $S$ stands for isoscalar-scalar, $V$ for isoscalar-vector, and $TV$ for isovector-vector type of coupling characterized by their transformation properties in isospin and in spacetime.
  
 The equations of motion for the nucleons and for the hyperon are obtained with the variational principle,  
 \begin{eqnarray}
 \label{Dirac}
 && \delta \langle \Phi^{(N\Lambda)}_{n}(\beta_2, \beta_3)\vert \hat H - \sum_{\tau=n, p} \lambda_\tau \hat N_\tau \nonumber\\
 &&- \sum_{\lambda=1, 2,3} C_\lambda (\hat Q_{\lambda0} - q_\lambda)^2  \vert \Phi^{(N\Lambda)}_{n}(\beta_2, \beta_3)\rangle=0
\end{eqnarray}
 with the Lagrange multipliers $\lambda_\tau$ being determined by the constraints $\langle q \vert \hat N_\tau\vert q\rangle=N(Z)$.  $C_\lambda$ is the so-called stiffness parameter. The position of the center of mass coordinate is fixed at the origin to decouple the spurious states by the constraint $\langle \Phi^{(N\Lambda)}_{n} \vert \hat Q_{10} \vert \Phi^{(N\Lambda)}_{n}\rangle=0$. The $\hat N_\tau$ and $\hat Q_{\lambda 0}\equiv r^\lambda Y_{\lambda 0}$ are particle number and multipole moment operators. The deformation parameters $\beta_\lambda$ ($\lambda=2, 3$) are defined as
\begin{equation}\label{deformation}
  \beta_\lambda \equiv \dfrac{4\pi}{3A R^\lambda} \langle\Phi^{(N\Lambda)}_{n}(\beta_2, \beta_3) \vert \hat Q_{\lambda0} \vert \Phi^{(N\Lambda)}_{n}(\beta_2, \beta_3) \rangle,
\end{equation}
with $A$ representing the mass number of the nucleus,  $\quad R=1.2A^{1/3}$.  The wave function for the whole $\Lambda$ hypernucleus can be written as
\beq
|\Phi^{(N\Lambda)}_{n}(\beta_2, \beta_3)\rangle
= |\Phi^N(\beta_2, \beta_3)\rangle \otimes |\varphi^{\Lambda}_{n}(\beta_2, \beta_3)\rangle,
\eeq
where $|\Phi^N(\beta_2, \beta_3)\rangle $ and
$|\varphi^{\Lambda}_{n}(\beta_2, \beta_3)\rangle$ are the mean-field wave
functions for the nuclear core and the hyperon, respectively. The index $n$ labels different $\Lambda$ hyperon states.
In this work, the $\Lambda$ is put on one of the two lowest-energy states, and is labeled as $\Lambda_s$ and $\Lambda_p$, respectively,
even though the $\Lambda$ wave function is generally an admixture of states with different orbital angular momenta due to the 
nonzero quadrupole-octupole deformation. Pairing correlations between nucleons are treated in the BCS approximation with a density-independent 
zero-range force  \cite{Yao10}. The pairing energy can be written as

 \beq
 \label{PairingDFT}
 E_{\rm pair}[\kappa,\kappa^*]
 = - \sum_{\tau=n,p} \dfrac{V_\tau}{4}\int d^3r\kappa^\ast_\tau(\bm{r}) \kappa_\tau(\bm{r}).
 \eeq
 where $V_\tau$ is a constant pairing strength, and the
 pairing tensor $\kappa(\bm{r})$ reads
 \beq
  \kappa(\bm{r})
  =-2\sum_{k>0}f_ku_kv_k\vert\psi_k(\bm{r})\vert^2,
 \eeq
with an energy-dependent regulator $f_k$ to  avoid divergence, 
 \beq%
 \label{Weight}
 f_k  =\frac{1}{1+\exp[( \epsilon_k- \lambda_\tau-\Delta E_\tau)/\mu_\tau]} \;.
\eeq%
 The $\epsilon_k$ is the energy of the $k$-th single-particle state. The parameters $\Delta E_\tau$
 and $\mu_\tau=\Delta E_\tau/10$ are chosen in such a way that
 $2\displaystyle\sum_{k>0}f_k = N_\tau +1.65N^{2/3}_\tau$, where
 $N_\tau$ is the particle number of neutrons or protons.

\subsection{The quantum-number projection for $\Lambda$ hypernuclei}

The mean-field wave function $|\Phi^{(N\Lambda)}_{n}(\beta_2,\beta_3)\rangle$ generated with the above
deformation constrained RMF calculations usually does not conserve parity, particle number and angular momentum.
The symmetry-conserved hypernuclear wave function can be constructed as
\beq
\label{porjection:wf}
|NZJK^\pi, n\rangle =  \hat{P}^J_{MK}\hat{P}^N \hat{P}^Z\hat P^\pi |\Phi^{(N\Lambda)}_{n}(\beta_2,\beta_3)\rangle,
\eeq
where the $\pi$ is a label for the parity of the state. The $\hat P^\pi$ is the parity projection operator defined by parity operator $\hat P$ as follows
\begin{equation}
\hat P^\pi = \dfrac{1}{2}(1+\pi\hat P).
\end{equation}
The particle-number projection operator is 
\beq
\hat P^{\tau} = \dfrac{1}{2\pi}\int^{2\pi}_0 d\varphi_{\tau}  e^{i(\hat
N_\tau-N_\tau)\varphi_{\tau}} \,, 
\eeq
with $\hat N_\tau$ the particle-number operator for either neutrons ($\tau=n$)
or protons ($\tau=p$). The angular-momentum projection operator is
\beq
\hat P^{J}_{MK}=\dfrac{2J+1}{8\pi^2}\int d\Omega D^{J\ast}_{MK}(\Omega) \hat
R(\Omega) \,,
\eeq
with $D^{J}_{MK}(\Omega)$ as Wigner-D function. The projector $\hat P^J_{MK}$ 
extracts from the intrinsic state  $|\Phi^{(N\Lambda)}_{n}(\beta_2,\beta_3)\rangle$ the component whose
angular momentum along the intrinsic $z$ axis is given by $K$.  
For simplicity, axial symmetry is imposed. In this case, there is no $K$ mixing, and its 
value is determined by third component of the angular momentum of the $\Lambda$ hyperon in this work.

The energy of each projected state $|NZJK^\pi, n\rangle$ is determined as
\beq
\label{energy-projection}
E^{JK\pi, n} = \dfrac{\langle NZ JK^\pi, n\vert H\vert NZJK^\pi, n\rangle}{\langle NZ JK^\pi, n \vert NZJK^\pi, n\rangle}.
\eeq

More details about the implementation of angular momentum projection into 
the RMF approach can be found in Refs. \cite{Yao09,Yao14,Mei16R}.

 \section{Results and discussions}
 \label{Sec.III} 
  
The variation of the total energy with respect to the single-particle wave function $\psi^B_k$ for nucleons or hyperon leads to two separate Dirac equations, which  are solved in a spherical harmonic oscillator basis within 12 major shells. The oscillator frequency is given by $\hbar\omega = 41A^{-1/3}$ (MeV).  Among all the relativistic point-coupling EDFs, only the 
 PC-F1 \cite{Burvenich2002_PRC65-044308} for the $NN$ interaction has been adopted to adjust $N\Lambda$ interactions. Here, we thus adopt the PC-F1  and the PCY-S2 (for the $N\Lambda$)  \cite{Tanimura2012_PRC85-014306} throughout the calculation.

 \begin{figure}[]
\centering
\includegraphics[width=7cm]{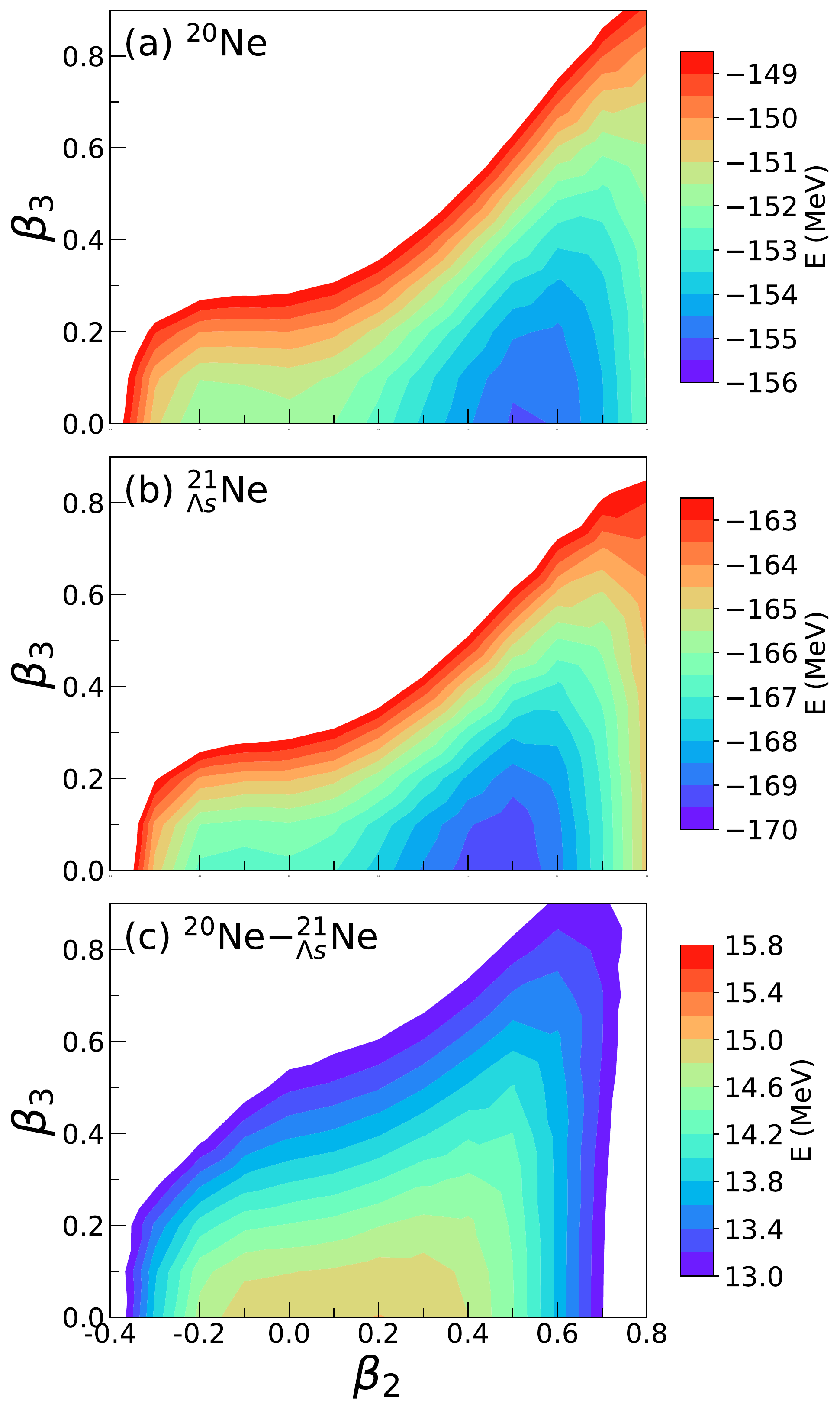} 
\caption{(Color online)  The contour plot of the energies for (a) $^{20}$Ne and (b) $^{21}_{\Lambda s}$Ne, as well as  (c) their difference,  in the ( $\beta_2, \beta_3$) plane. }
\label{PCF1:20Ne-contour}
\end{figure}
 
 Figure~\ref{PCF1:20Ne-contour} displays the contour plot of the total energy for $^{21}_{\Lambda s}$Ne and its core nucleus $^{20}$Ne in the quadrupole-octupole deformation ( $\beta_2, \beta_3$)  plane. The $\Lambda$ binding energy defined as the energy difference between them  is also shown.  The topology of the energy surface for $^{20}$Ne is similar 
 to that published in Ref. ~\cite{Zhou16}, even though the PC-F1 \cite{Burvenich2002_PRC65-044308}, instead of the PC-PK1 ~\cite{Zhao10} parameterization of the $NN$ effective interaction is used here.   It is shown that the energy surface is soft against the octupole shape fluctuation around the quadruple deformation $\beta_2$ in between 0.5 and 0.6.  
 With the presence of a $\Lambda_s$ hyperon, the softness along the octupole deformation is quenched. It can be seen clearly from the distribution of the $\Lambda_s$ binding energy, c.f. Fig.~\ref{PCF1:20Ne-contour}(c).  The $\Lambda$ binding energy decreases significantly along the optimal path of the total energy surface of  $^{20}$Ne, where the molecular cluster structure $^{16}$O$+\alpha$ is progressively developing. Of particular interest is the behavior of the $\Lambda_s$ binding energy as a function of the octupole deformation for a fixed quadruple deformation. In the small deformation region,  the  $\Lambda_s$ binding energy decreases rapidly with $\beta_3$, but is almost constant with the $\beta_2$. In the large deformation region, an opposite phenomenon is observed.  
 
  \begin{figure}[]
\centering
\includegraphics[width=7cm]{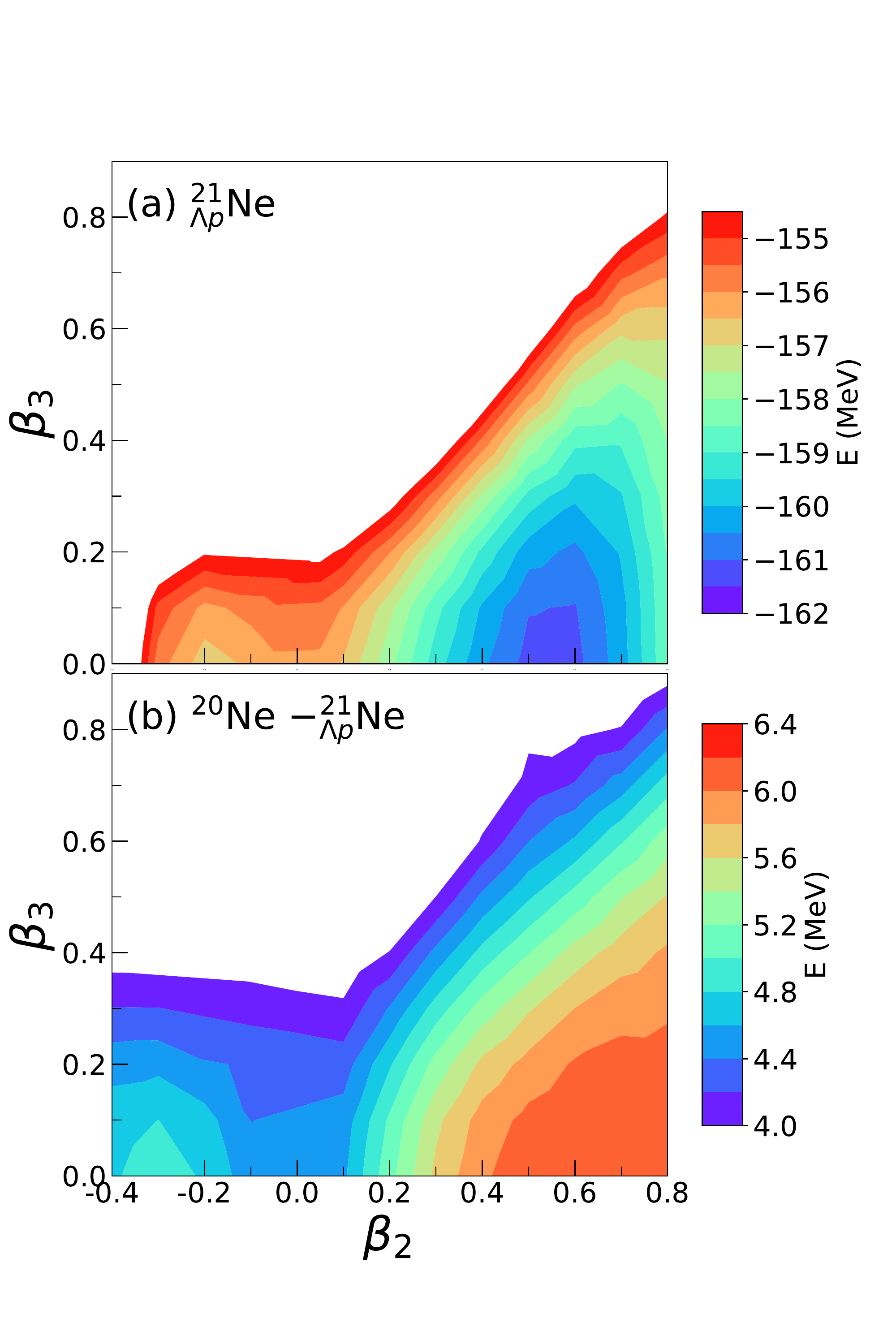} 
\caption{(Color online) The contour plot of (a) the energy for  $^{21}_{\Lambda p}$Ne and (b)  that of the energy difference between
 $^{20}$Ne and $^{21}_{\Lambda p}$Ne  in the ( $\beta_2, \beta_3$) plane. }
\label{PCF1:21pNe-contour}
\end{figure}

\begin{figure}[]
\centering
\includegraphics[width=8.5cm]{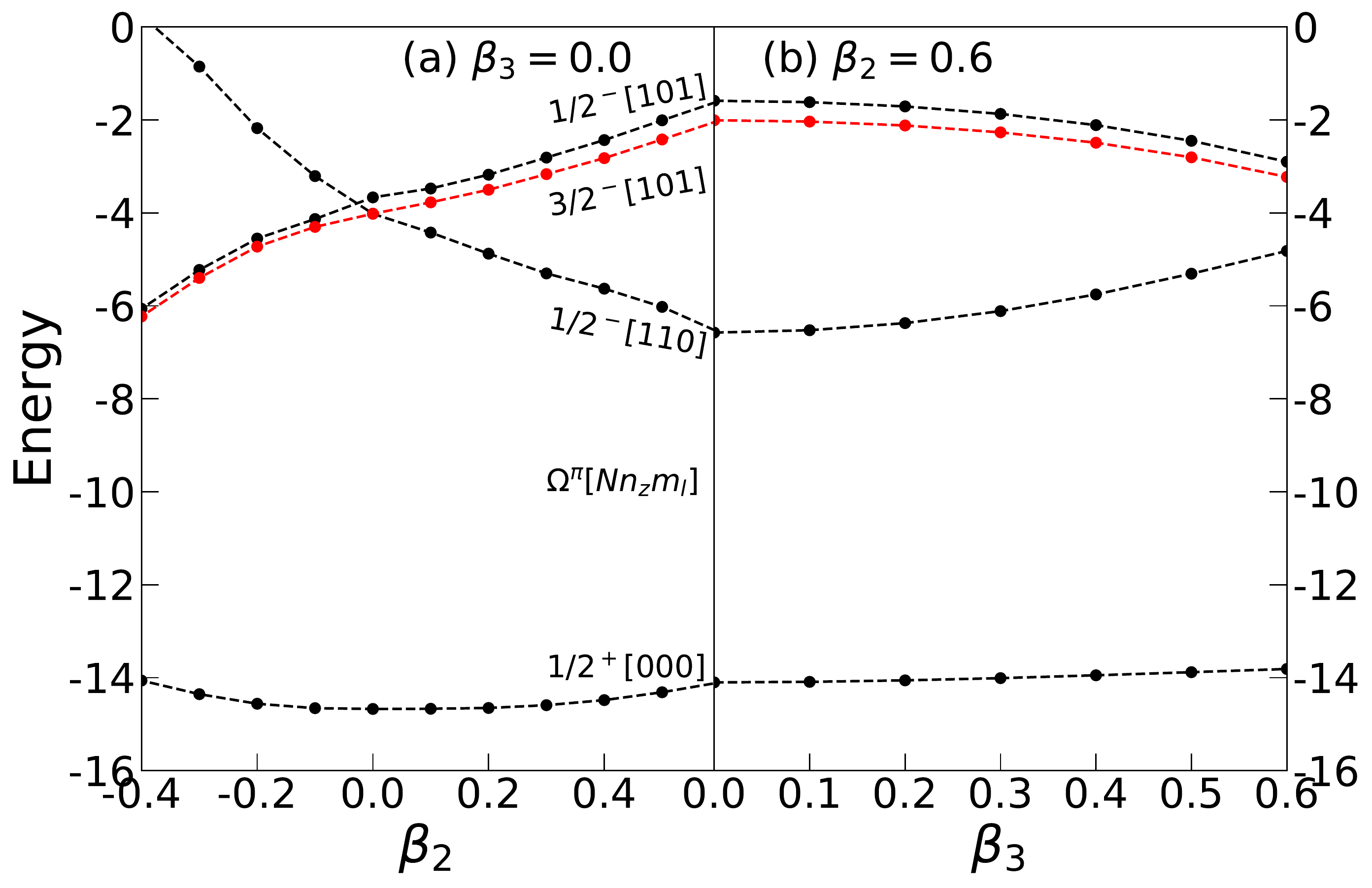}
\caption{(Color online)  (a) The Nilsson diagram for the $\Lambda$ hyperon in $^{21}_\Lambda$Ne from the RMF calculation as a function of quadrupole deformation $\beta_2$ at $\beta_3=0$ and (b) that as a function of octuple deformation $\beta_3$ at $\beta_2=0.6$. The levels in (a) are labeled with Nilsson quantum numbers $\Omega^\pi[Nn_zm_\ell]$,
where $\Omega^\pi$ ($\Omega$ is the third component of angular momentum, $\pi$ is parity) are good quantum numbers for reflection-symmetric axially deformed states.}
\label{fig:20Ne-spe}
\end{figure}

 In contrast, the single-particle energy of the second lowest-energy state is changing dramatically with both the quadrupole and octupole deformation parameters. Fig.~\ref{PCF1:21pNe-contour}
 displays the contour plot of the energy surface of $^{21}_{\Lambda p}$Ne in the $\beta_2$-$\beta_3$ plane, where the $\Lambda$ is put on the second lowest-energy state.
 It is seen that the energy minimum of  $^{21}_{\Lambda p}$Ne is more pronounced than that of  both  $^{20}$Ne and  $^{21}_{\Lambda s}$Ne. In other words, the $\Lambda_p$ favors the hypernucleus to be reflection-symmetric prolate shape.  It should be pointed out that the parity is violated in the configuration with nonzero $\beta_3$ value, but the angular momentum projection $\Omega$ is still conserved in this study. Since the $\Lambda p$ labels the hyperon in the second lowest-energy state, the $\Omega$ value in the configuration with $\beta_2<0$ is shifted to $3/2$, c.f. Fig.~\ref{fig:20Ne-spe}.

The change of $\Lambda$ binding energy against deformation is dominated by the behavior of single-particle energy. To understand the hyperon impurity effect, the Nilsson diagram for the $\Lambda$ hyperon in $^{21}_\Lambda$Ne is plotted in Fig.~\ref{fig:20Ne-spe}. One can see that the lowest-energy level which in reflection-symmetric case is labeled as $1/2^+[000]$ is somewhat raising up with the increase of $\beta_2$ up to 0.6, at which quadrupole deformation, its energy is almost constant with the increase of $\beta_3$.  The evolution trend with deformation parameters is consistent with that shown in the $\Lambda$ binding energy in Fig.~\ref{PCF1:20Ne-contour}(c).

  \begin{figure}[h]
\centering
\includegraphics[width=9.5cm]{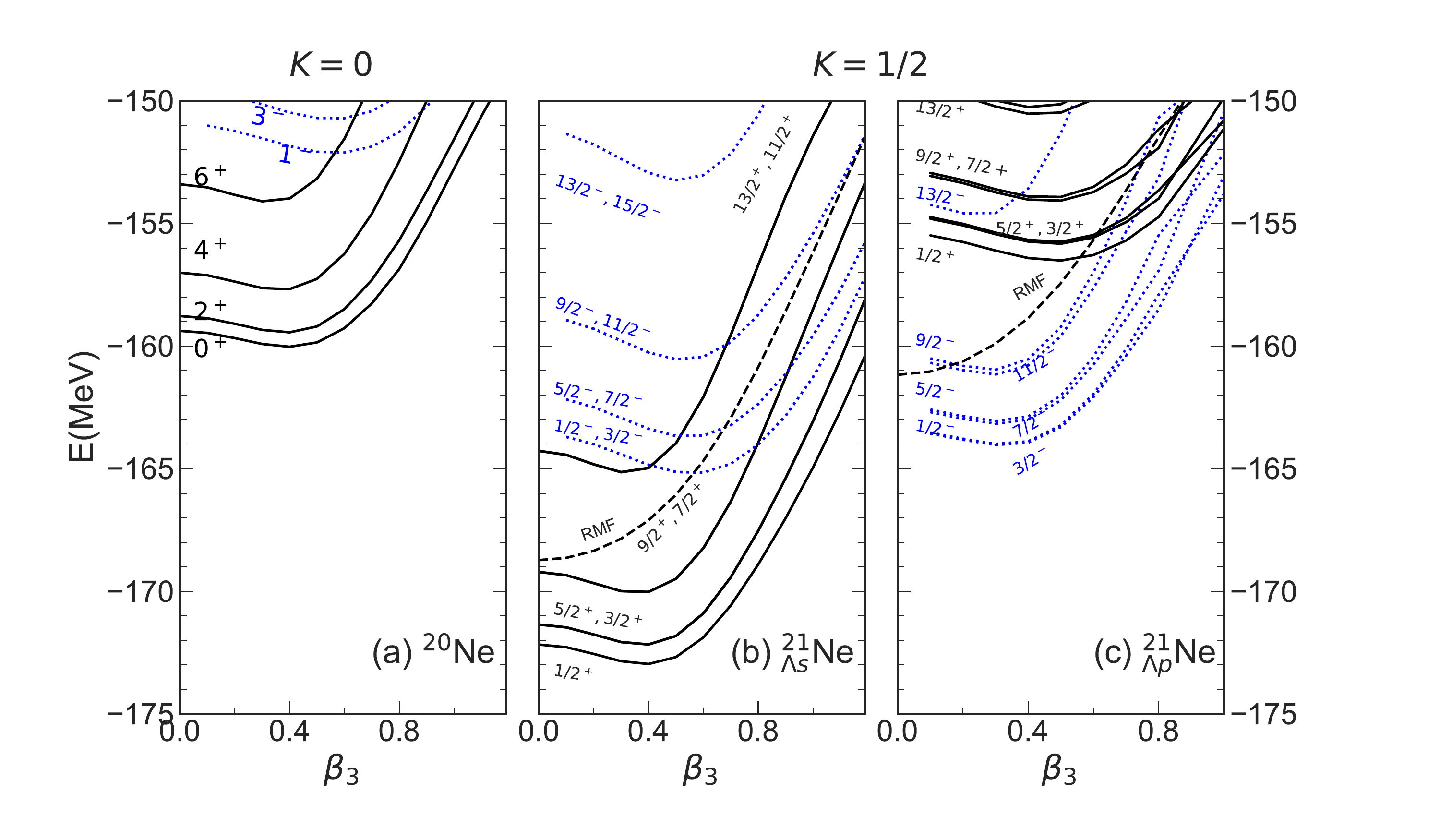}
\caption{(Color online)   The projected energies of levels with spin-parity $J^\pi$ in  (a) $^{20}$Ne, (b) $^{21}_{\Lambda s}$Ne, and (c) $^{21}_{\Lambda p}$Ne as a function of the octupole deformation $\beta_3$ of the  intrinsic state ($\beta_2=0.60$ is fixed).  The levels with positive or negative parity are plotted solid or dotted curves, respectively.}
 \label{fig:21Ne-projection}
\end{figure}

  \begin{figure}[]
\centering
\includegraphics[width=4cm]{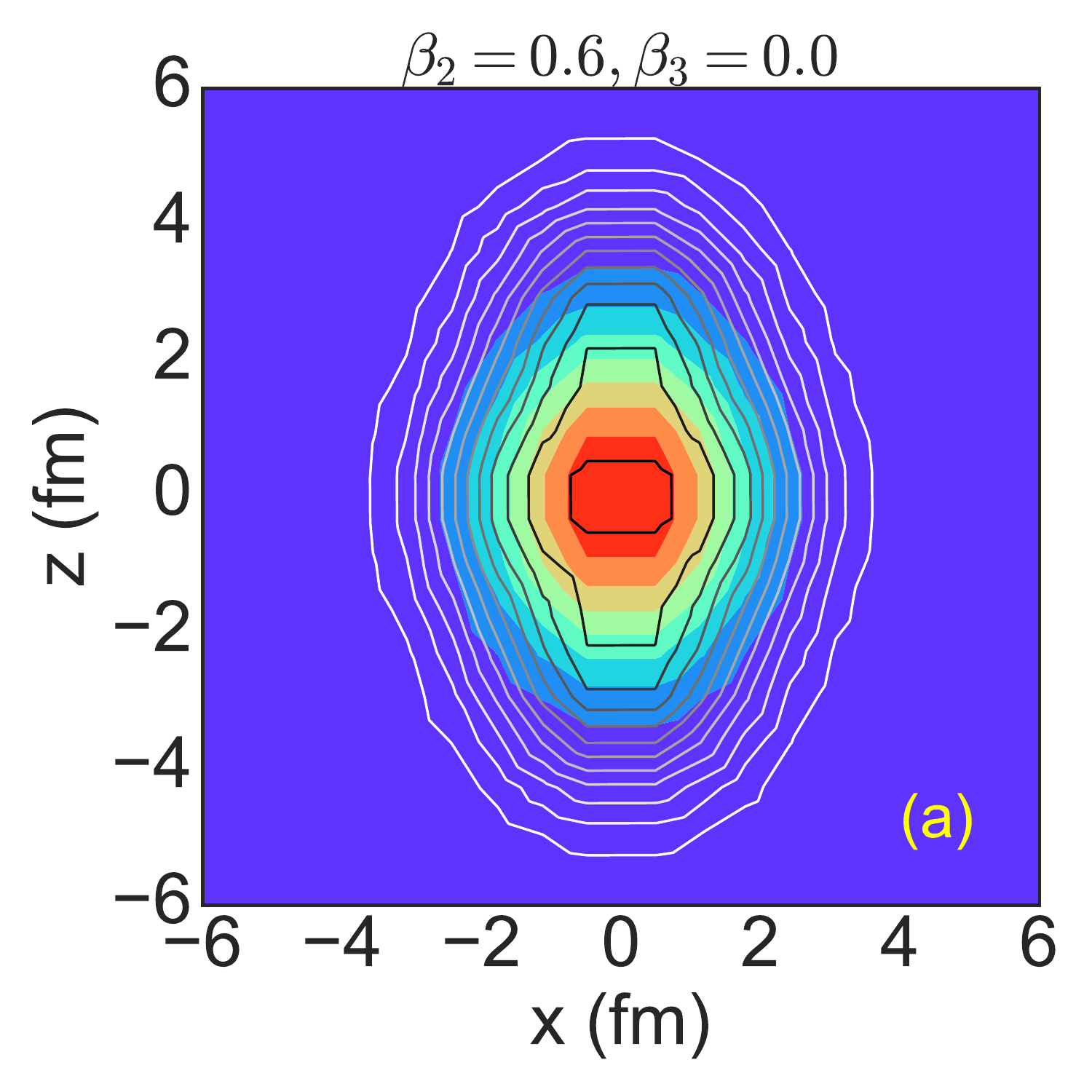} 
\includegraphics[width=4cm]{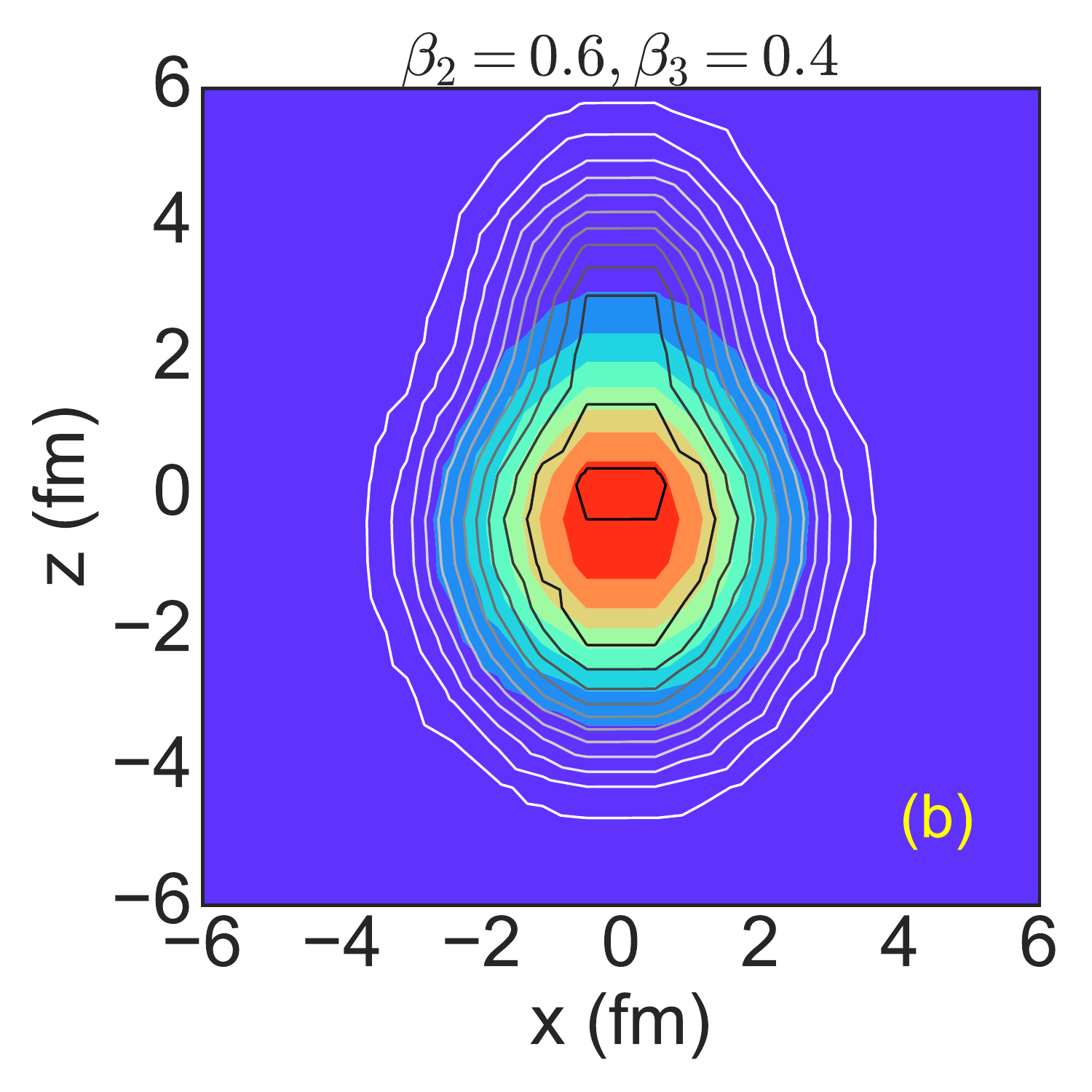} 
\includegraphics[width=5cm]{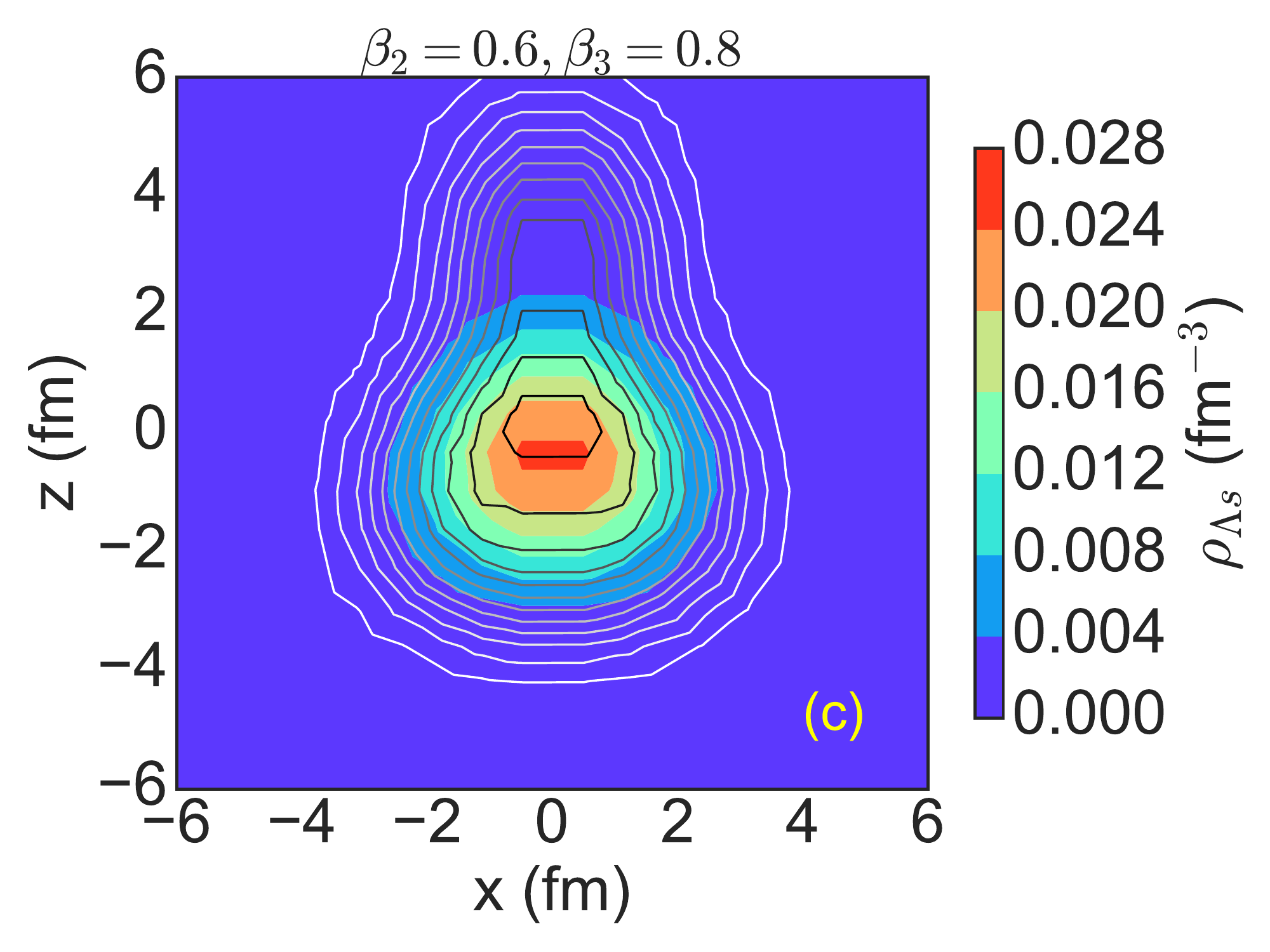} 
\caption{(Color online) The contour plot of the nucleon density  (difference between two neighboring lines is 0.015 fm$^{-3}$) and the density profile of the $\Lambda$   in the ($x, z$) plane at $y=0$ fm for several different intrinsic states of $^{21}_{\Lambda s}$Ne with $\beta_2=0.6$ and $\beta_3=0.0, 0.4, 0.8$, respectively. }
\label{PCF1:21sNe-dens}
\end{figure}

  \begin{figure}[]
\centering
\includegraphics[width=4cm]{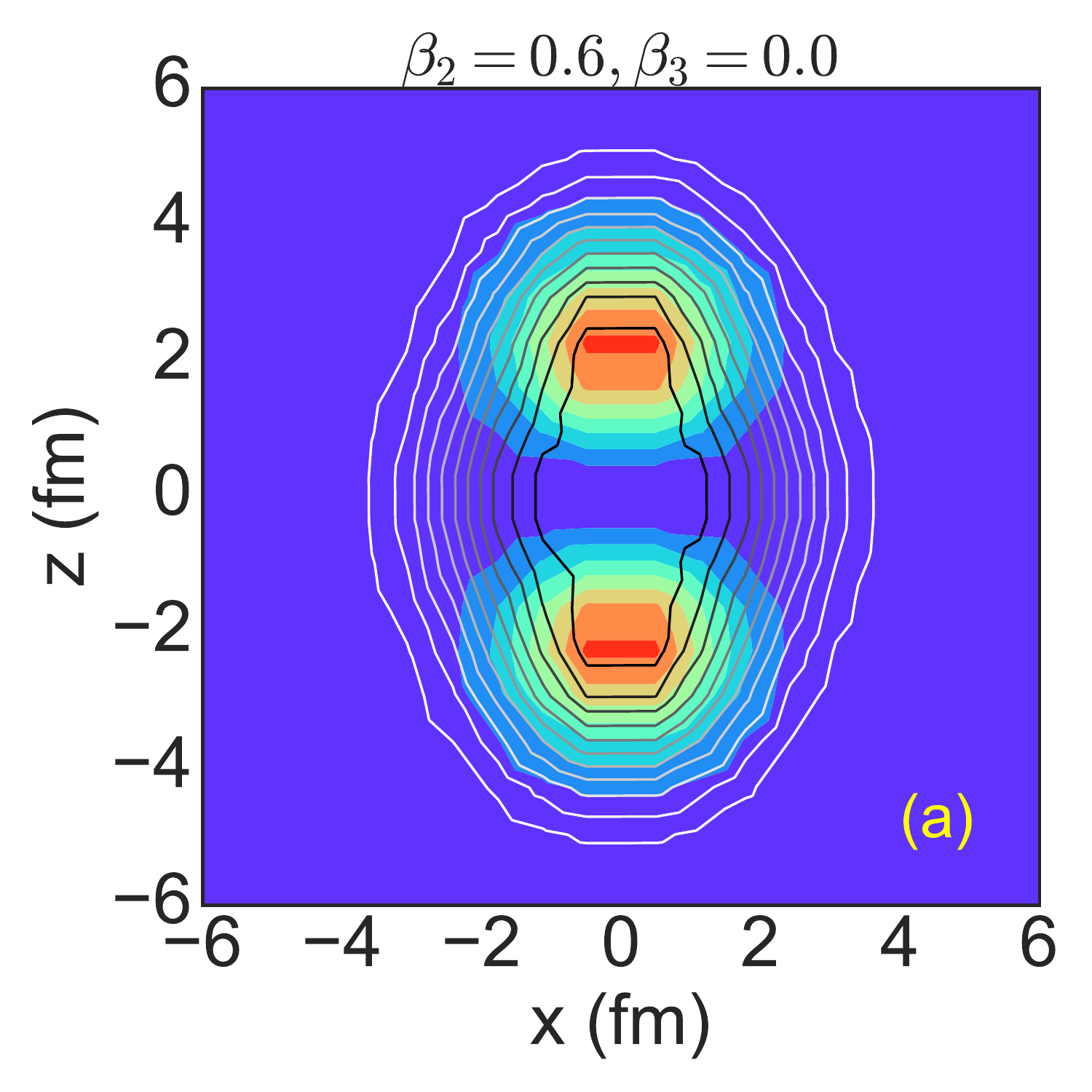} 
\includegraphics[width=4cm]{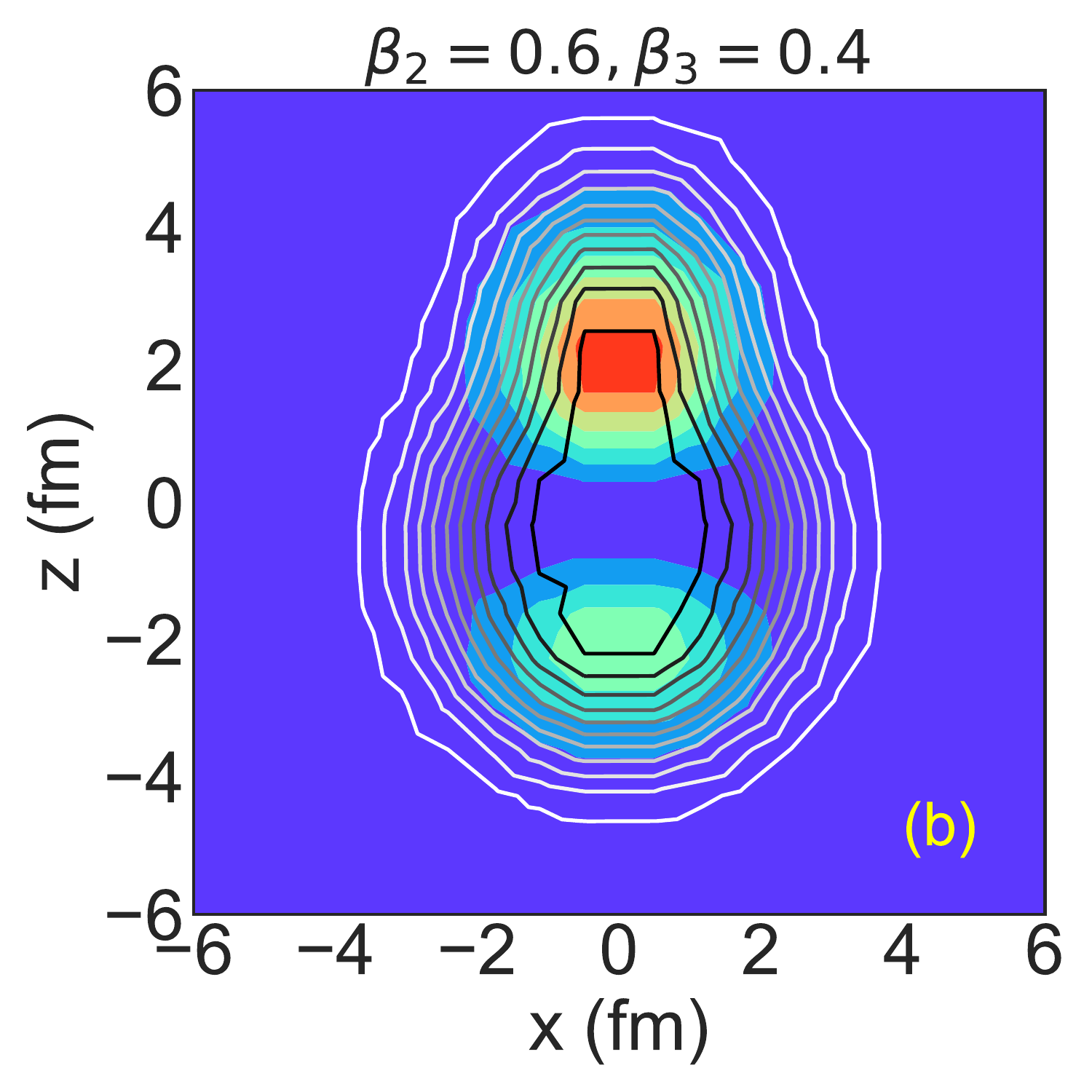} 
\includegraphics[width=5cm]{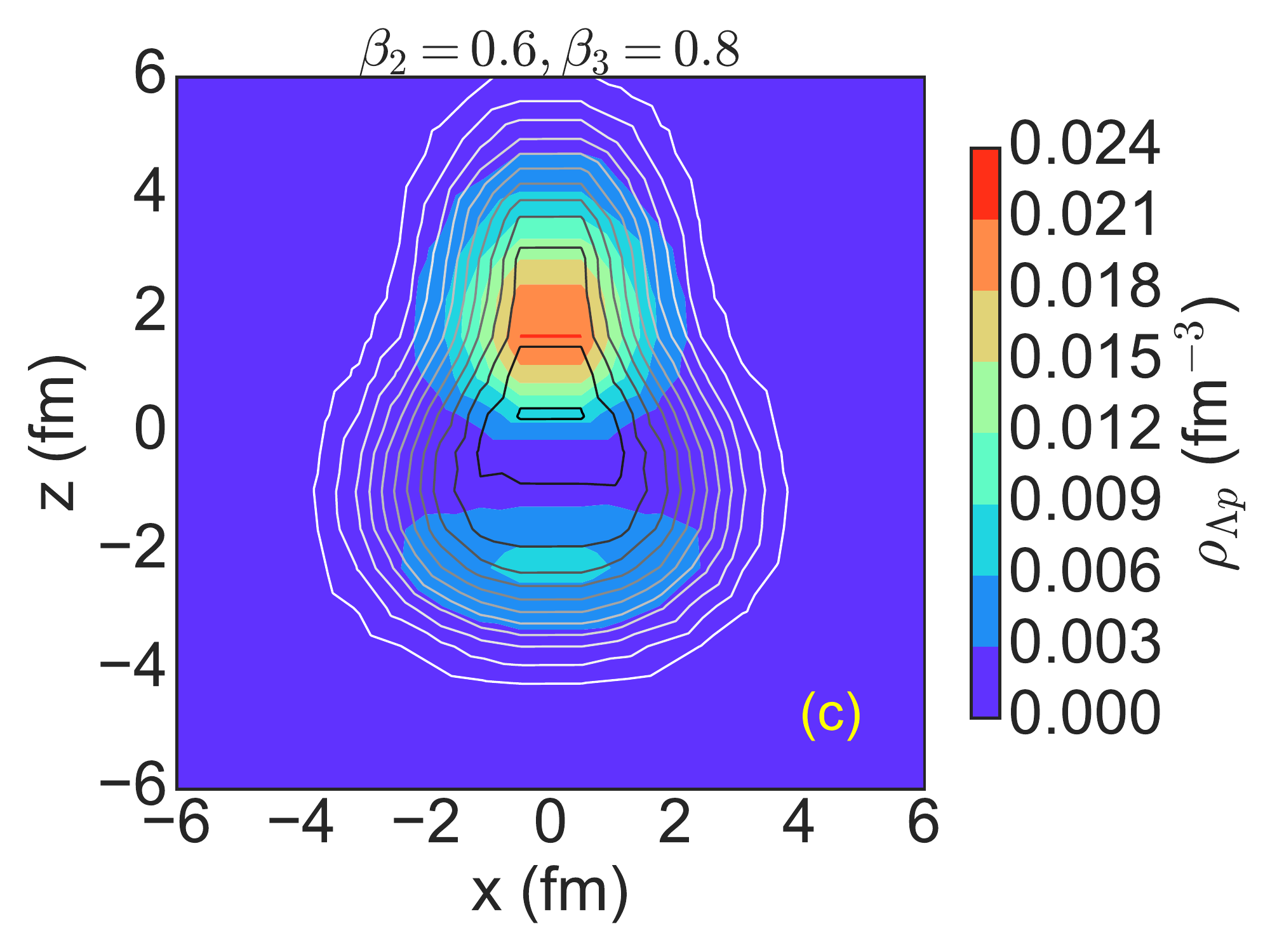} 
\caption{(Color online) The same as Fig. \ref{PCF1:21sNe-dens}, but for the $\Lambda$ hyperon on the $\Lambda_p$ state. }
\label{PCF1:21pNe-dens-p}
\end{figure}

The occurrence of softness along octupole shape degree of freedom is accompanying with the existence of low-lying parity-doublet states. A similar phenomenon is expected in  $\Lambda$ hypernuclei. The hypernuclear AMD model has been adopted to study the low-lying parity-doublet states in $^{21}_\Lambda$Ne~\cite{Isaka11} with the configuration of
$^{20}$Ne($K^\pi=0^+)\otimes \Lambda_s$ and $^{20}$Ne($K^\pi=0^-)\otimes \Lambda_p$. It has been found there that the  reduction in the intraband electric quadrupole (E2) transition stengths in the $^{20}$Ne($K^\pi=0^-)\otimes \Lambda_p$ band  is larger than that in the  $^{20}$Ne($K^\pi=0^+)\otimes \Lambda_s$ band, which was attributed to the reduction of the intercluster distance between $\alpha$ and $^{16}$O clusters in the band of $^{20}$Ne($K^\pi=0^-)$. 

Here, we are more interested in the evolution of the energies $E^{JK\pi, n}$, defined in Eq.(\ref{energy-projection}), of the low-lying states in $^{21}_\Lambda$Ne with the quadrupole-octupole deformation, as depicted in Fig.~\ref{fig:21Ne-projection} , where the $\Lambda$ is put in one of the two lowest-energy states. The quadrupole deformation is fixed at $\beta_2=0.6$, around its energy minimum. The energy levels display the patterns of $1/2^+, (5/2^+, 3/2^+), \ldots$ and $(3/2^-, 1/2^-), (7/2^-, 5/2^-), \ldots$, regardless of the values of the quadrupole-octuple deformation of concerned. The energy splitting in the spin doublets is negligible in $^{21}_{\Lambda s}$Ne, but is increasing with $\beta_3$ in $^{21}_{\Lambda p}$Ne. It is interesting to note that the location of 
the negative-parity states with the configuration $^{20}$Ne($K^\pi=0^+)\otimes \Lambda_s$ is close to that with the configuration $^{20}$Ne($K^\pi=0^-)\otimes \Lambda_p$, even though they have very different structures. It is difficult to distinguish them from each other experimentally simply based on energies. We note that the E2 transition strengths in the $K^\pi=0^-$  band of $^{20}$ is about three times of those in the $K^\pi=0^+$  band. It is reasonably  to anticipate that this relation holds in $^{21}_\Lambda$Ne.  Moreover, the configuration mixing is not considered in this work. We leave all these studies for the future.

The density distribution of nucleons and that of the hyperon in $^{21}_{\Lambda s}$Ne and $^{21}_{\Lambda p}$Ne are displayed in Figs.  \ref{PCF1:21sNe-dens} and \ref{PCF1:21pNe-dens-p} respectively. The contour lines represent the density of nucleons. With the increase of octupole deformation, the pear-like shape is progressively developed as expected. Of particular interest is the change of the distribution of the hyperon. For the $\Lambda_s$, the density is more and more concentrated around the bottom of the ``pear". This phenomenon is consistent with the finding in the hypernuclear AMD study \cite{Isaka11}.  The new finding here is the change of the density for the $\Lambda_p$, which is not shown in Ref. \cite{Isaka11}.  It is seen that the $\Lambda_p$ becomes more and more concentrated around the top of the ``pear", which is very interesting. 
The density distribution of the $\Lambda$ shows clearly that the octuple shape violates the parity of the $\Lambda_s$ and $\Lambda_p$, both of which are actually admixtures of positive and negative-parity orbits.

%
\section{summary}
\label{Sec.IV}
%
%
  
We have developed both relativistic mean field and beyond approaches for hypernuclei with possible  pear-like shapes based on relativistic point-coupling energy density functionals.  The techniques of parity, particle-number and angular momentum projections have been implemented to restore the symmetries broken in the mean-field configurations. The method has been illustrated by taking $^{21}_\Lambda$Ne as an example, where the $\Lambda$ is put on one of the two lowest-energy orbits, respectively.  We have found that the $\Lambda$ hyperon in both cases disfavors the formation of reflection-asymmetric molecular-like $^{16}$O$+\alpha$ structure in $^{20}$Ne. In particular, we have shown that the energies of the negative-parity states with the configuration $^{20}$Ne($K^\pi=0^-)\otimes \Lambda_s$ are close to those with the  configuration $^{20}$Ne($K^\pi=0^+)\otimes \Lambda_p$. Moreover, we have shown that the $\Lambda_s$ ($\Lambda_p$)  becomes more and more concentrated around the bottom (top) of the ``pear" with the increase of octupole deformation.
It is interesting to go a step further to perform a configuration mixing calculation  for the low-lying states in $^{21}_\Lambda$Ne within the generator coordinate method.
Work along this direction is in progress. Moreover, a beyond RMF calculation of hypernuclei with other exotic shape degrees of freedom, such as tetrahedral shapes \cite{Zhao17}, is also very interesting.

\begin{acknowledgements}
 Helpful discussions with K. Hagino, B. N. Lu, T. Motoba are acknowledged.  This publication is based on work supported in part by National Natural Science Foundation of China under Grant No. 11575148. JMY also acknowledges the support by the Scientific Discovery through Advanced Computing (SciDAC) program funded by the U.S. Department of Energy, Office of Science, Office of Advanced Scientific Computing Research and Office of Nuclear Physics, under Award Number DE-SC0008641 (NUCLEI SciDAC Collaboration).
\end{acknowledgements}



%

\end{CJK*}
\end{document}